\documentstyle[epsf,prl,aps,twocolumn]{revtex}
\begin{document}
\draft
\title{How loud can   Schwarzschild black holes ring? }
\author{Janusz Karkowski, Edward Malec}
\address{Institute of Physics, Jagiellonian University,
 30-059 Krak\'ow, Reymonta 4, Poland}
\author{and Zdobys\l aw \'Swierczy\'nski}
\address{Department of Computer Science and Computer Methods,
Pedagogical University of Cracow,
    Krak\'ow,  Podchor\c a\.zych 1, Poland}

\maketitle

\begin{abstract}
A numerical procedure is described for the maximization of the energy
diffusion due to the backscattering of the gravitational radiation. The
obtained maxima are solutions dominated by low frequency waves.
They  give rise to robust gravitational ringing, with   amplitudes 
of the order of the original signal.  
\end{abstract}

\date{today }

\section{Motivation}

The detection of gravitational radiation by the first generation of
laser interferometers requires from theorists
and numericists developing of the so-called templates for the gravitational
waves (GW) produced by binary black holes (BH) \cite{Thorne},
\cite{Damour}. That can be done for
the inspiral of BH while there are significant problems as it concerns  the
merger  phase  of BH \cite{Thorne99}, where no numerical data are available as yet.
In this context the   next,    ringing phase,
of the collapse of binary BH is interesting because  GW  
 produced during this evolution period
are   independent to a degree on the details of its preceding history.
Numerical simulations demonstrate that irrespective of the initial
data, during some "intermediate" time   the outgoing  GW
are dominated by a (fundamental) quasinormal mode (QM) \cite{all}, whose oscillation
period and the damping coefficient depend only on characteristics of the final
black hole.   It is  not clear, however, how energetic 
\cite{Nollert99} are the QM  and how big is the ringing amplitude. The full
answer to this question would require  detailed information on the
former evolution of the system, which is not available at present.
This paper aims at finding "upper limits", that is identifying
the most favourable situations  for ringing and comparing their strength
- amplitudes and the  energy content - with  the initial energy
and amplitudes. The amplitudes of strongest ringing modes are  
of the order of the amplitudes of  GW that generate  them.

\section{Extremizing the diffused energy}

 {\it A. Prerequisities.}
The space-time geometry  is defined  by a line element,
\begin{equation} ds^2 = - (1-{2m\over R})dt^2 +
{1\over 1-{2m\over R}} dR^2 +
R^2 d\Omega^2~,
\label{1}
\end{equation}
where $t$ is a time coordinate, $R$ is  
 the areal radius and $d\Omega^2 = d\theta^2
+ \sin^2\theta d\phi^2$ is the line element on the unit sphere,
$0\le \phi < 2\pi $ and $0\le \theta \le \pi $. The Newtonian
gravitational constant $G$, and $c$, the velocity of light are put
equal to 1.  

  We will study the propagation of polar modes $\Psi $
of the quadrupole   GW in a Schwarzschild background;
they are ruled by the Zerilli equation \cite{Zerilli}
\begin{equation}
(-\partial_t^2 + \partial_{r^*}^2)\Psi  = V\Psi .
\label{2}
\end{equation}
Here    $\eta_R=1-2m/R$,  $V(R)=6\eta_R^2  {1\over R^2} +
\eta_R{63m^2(1+{m\over R})\over 2R^4(1+{3m\over 2R})^2}$
and  $r^*=R+2m\ln ({R\over 2m}-1)$.

There exists a positive  conserved  energy  $E$ \cite{energy}; its quasilocal
contribution coming from a region of a Cauchy hypersurface $\Sigma_t$ exterior
to a sphere of radius $R$ reads
\begin{equation}
E(R,t) =2\pi  \int_{R}^{\infty }dr
\rho (r,t) .
\label{3}
\end{equation}
Here  $\rho $ is the reduced energy density,
$\rho =\bigl( (\partial_t\Psi )^2 + (\partial_{r^*}\Psi )^2
+V\Psi^2\bigr) /\eta_R $.
The initial data  $\Psi  $ and $\partial_t\Psi  $
are assumed to be purely outgoing, smooth and to be nonzero
outside a sphere of a radius $a >2m$. Thus
$\rho $ is  smooth and  vanishes inside the sphere of radius $a$.

Let   $\tilde \Gamma_{(R,t)} $ be  an outgoing null geodesic
that  originates at $(R,t)$. By $\tilde \Gamma_{(R_0, t_0), (R,t)}$
will be understood a segment of $\tilde \Gamma_{(R_0, t_0)}$
ending at $(R,t)$. A straightforward calculation shows that
the rate  of the energy change along $\tilde \Gamma_{(R,0)}$ is
given by
\begin{eqnarray}
&&(\partial_t+\partial_{r^*})E(R,t)= \nonumber\\
&&- 2\pi \Biggl[ \Bigl(\partial_t\Psi +\partial_{r^*}\Psi \Bigr)^2
  +V \Psi^2 \Biggl] .
\label{4}
\end{eqnarray}
Thus $E(R,t)=E(a,0)-2\pi \int_0^tdt \Biggl[ \Bigl(\partial_t\Psi +
\partial_{r^*}\Psi \Bigr)^2  +V \Psi^2 \Biggl] $, where the integral
is done along $\tilde \Gamma_{(a, 0), (R,t)}$.  Some of the energy
can diffuse inward due to the scattering off the curvature of the spacetime
\cite{back}. In the limiting case  case when  the integration  contour
coincides with $  \tilde \Gamma_{(a,0)}$, 
the energy limit  $E_B(a,0)=\lim_{t\rightarrow \infty } E(R(t), t)$
is    the analogue of the Bondi mass.
We will define the diffused energy as $\delta E_a=E(a,0)-E_B(a,0)$.
Our aim is to find initial data, that maximize the {\bf diffusion
parameter}  $\kappa \equiv \delta E_a/E(a,0)$ for a selection of {\it a}'s.
Let us remark that, from our experience, 
the computational time is proportional to the 
square $(a/(2m))^2$; therefore  the maximization
procedure would not be feasible  in the case $a>>2m$ .
Fortunately, it is known from analytic estimates, that for 
big $ a$'s the diffusion is negligible, since $\kappa <C (2m/a)^2$  
(where $C$ is of the order of 10)
\cite{Gerhard}, \cite{kark2001}.
 Therefore it suffices to focus to the range of
relatively small distances - $a$ being of the order of the Schwarzschild
radius $2m$ -  and in that case the numerical methods appear to
be efficient.

 {\it B. Method.}
It is advantageous to   search a solution of Eq. (\ref{2}) in the form 
  $\Psi =\tilde \Psi +\delta  $ \cite{Gerhard}.
Here $\tilde \Psi $ is a known function that is constructed from initial
data while $\delta $ is the  unknown part of the sought solution.
In explicit terms
\begin{equation}
\tilde \Psi \equiv  \Psi_0(r^*-t)+ {  \Psi_1(r^*-t)\over R} +
{  \Psi_2(r^*-t)\over R^2},
\label{8}
\end{equation}
where $ \Psi_i(r^*-t)$, $i=0,1,2$ fulfill the  relations
$\partial_t  \Psi_1= 3  \Psi_0$ and $\partial_t   \Psi_2= \Psi_1 - 
m\partial_t \Psi_1$.
$\tilde \Psi $ solves the Zerilli equation in  Minkowski space-time ($m=0$),
in which case   it represents a purely outgoing radiation.
The function  $\delta $    satisfies the equation
\begin{eqnarray}
&&(-\partial_t^2 + \partial_{r^*}^2)\delta  = V\delta +
(V-6{\eta_R^2\over R^2})\Biggl( \Psi_0+{\Psi_1\over R}+{\Psi_2\over
R^2}\Biggr) +
\nonumber\\
&&  { 2m\eta_R\over R^4}\Biggl[ -3\Psi_1 +2{\Psi_2\over R}  \Biggr] ,
\label{10}
\end{eqnarray}
with initial data  $\delta =\partial_t\delta =0$ at $t=0$.

Eq. (\ref{8}) and the relations between functions $\Psi_i$
  imply   that initial data $\Psi $ and $\partial_t\Psi $
are entirely determined by the specification
of  $\Psi_0$. $\Psi_0 $ in turn is determined (in a region outside the cone
defined by $\tilde \Gamma_a $) by choosing  
$\partial_t\Psi_0 $ (with $\partial_t\Psi_0|_a=0$)
in an interval $(r^*(a),   \infty )$  
and  $\Psi_0=0$ for $R = a$.
We have found that the numerical calculations can be significantly 
facilitated if the following procedure is applied.
First, expand $\partial_t\Psi_0$ in an 
$n$-element Chebyshev polynomial basis $(f_i)$ \cite{Chebyshev};
 one has
\begin{equation}
\partial_t\Psi_0 ( r^*, t=0)=\Sigma_{i=1}^nC_if_i(r^*).
\label{5}
\end{equation}
The initial energy becomes in this basis  quadratic in $C_k$.
 $E(a,R,t=0)\equiv 2\pi \int_a^Rdr\rho (r, 0)=\Sigma_{i,k=1}^nB_{ik}(a,R)C_iC_k$.
Second, let $\Psi_{f_k}(r^*, t)$ be a solution of (\ref{2})  
with initial data being defined by $f_k(r^*)$. Obviously
$\Psi_{ f_k}(r^*,t)$ depends only on the initial data 
within  the null past cone with the apex at  $(r^*(a),t)$. 
Then the solution of Eq. (\ref{2}) reads
\begin{equation} 
\Psi (r^*,t)= \Sigma_{i=1}^nC_i\Psi_{ f_i}(r^*,t).
\label{5a}
\end{equation}
Let $R_1= R(r^*=t+r^*(a) )$
and $R_2=R(r^*=2t+r^*(a) )$. The   energy diffusion through
$\tilde \Gamma_{(a,0), (R_1, t)}$  is
 of the   form
\begin{equation}
\delta E_a(R,t)=  \Sigma_{i,k=1}^n  A_{ik}(a,R)C_iC_k.
\label{6}
\end{equation}
The quadratic form $\delta E_a(t, R_1)$ is nonnegative; since
$E(a, R,t=0)$ is positive, the two forms  
  can be simultaneously diagonalized. The task of
comparing the extremal ratio of $\delta E_a$ and $E(a,R, t=0)$
can be reduced to the search of  the maximal eigenvalue of the generalised
eigenvalue problem $AX=\lambda BX $. That has been  accomplished
with the use of  the fast {\it EISPACK}  \cite{netlib} package.
The typical step used in numerics was $\Delta =0.02$ and $n$ was of the order of
one thousand, depending (weakly) on the size of the initial support.

 {\it C. Extremal initial data. }
Below the mass $m$ has been normalized to unity.
The method described above consists in comparing local energy expressions -
the energy diffused through a segment  $\tilde \Gamma_{(a,0),(R_1, t)}$ is
compared with the energy content inside the region 
$a\le R\le R_2$  of the initial hypersuface. Fixing $a$ and 
$R_2$, one finds  $\partial_t\Psi_0^{R_2}$ (the upper index is put here in order
to stress the local character of the procedure)   and, through (\ref{8}),
initial values of the   locally extremizing solution $\Psi^{R_2}$.
With the increase of $R_2$, while  keeping $a$ fixed, the function
$\partial_t\Psi_0^{R_2}$ changes. In the limit   one obtains the sought 
extremizing solution,  $\Psi =\lim_{R_2\rightarrow \infty } \Psi^{R_2}$.
In the numerical  practice the integration region must be finite.

Fig. 1 shows the initial profile 
$\partial_t\Psi_0$ for $a=2.1$  and for various $R_2$.
\begin{figure}[1]
\epsfxsize=6cm
\centerline{\epsffile{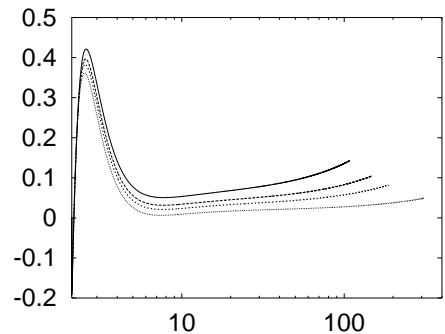}}
\caption{
  Function $\partial_t\Psi_0$ in dependence on  $R_2$. The abscissa ($R-$ axis)
is in the (decimal) logarithmic  scale. Here $a=2.1$; solid line,  long broken line and short broken line correspond to $r^*(R_2)=r^*(2.1)+6000$, $r^*(R_2)=r^*(2.1)+8000$ and  $r^*(R_2)=r^*(2.1)+10000$, respectively. Dotted line depicts the case with  $r^*(R_2)=r^*(2.1)+16000$.}
\end{figure}

The dependence of $\partial_t\Psi_0$ on $R_2$ 
suggests that  $\lim_{R_2\rightarrow \infty }  \partial_t\Psi_0^{R_2}\approx 0$
outside some region of compact support.
The parameter $\kappa $ was extremized in that class of  initial data
which is characterized by the vanishing of $\partial_t\Psi_0$ outside 
some bounded region.
Therefore the chosen $\Psi_0$   bears on an asymptotically constant value
 \cite{remark}.   Fig. 2 shows   initial profiles of
$\partial_t\Psi_0$  for $a=2.1$, $a=3.1$ and $a=4$.

The obtained extremizing initial data have a  {\bf finite }
total energy.  Fig. 3 shows initial energy  profiles   for $a=2.1$,
 $a=3.1$ and $a=4$. It is interesting to notice how efficient 
is the backscatter  - $\kappa $ ranges from c. 90\% (a=2.1)  and c. 16 \%
(a=3.1)  to c. 4\% (a=4).  Notice also that the region with the largest 
contribution to the initial energy  widens with the increase of $a$.
\begin{figure}[2]
\epsfxsize=6cm
\centerline{\epsffile{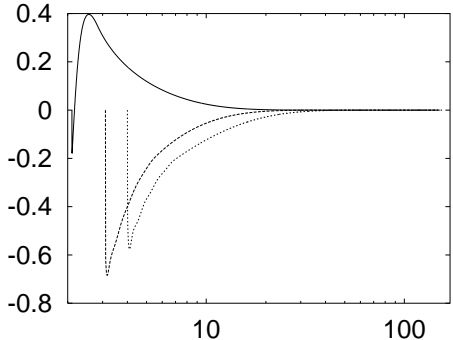}}
\caption{ Function $\partial_t\Psi_0$ for various values of $a$. The abscissa ($R-$ axis)
is in the (decimal) logarithmic
  scale. Solid line,  long broken line and short broken line correspond to $a=2.1$, $a=3.1$ and $a=4$, respectively.}
\end{figure}
 
\begin{figure}[3]
\epsfxsize=6cm
\centerline{\epsffile{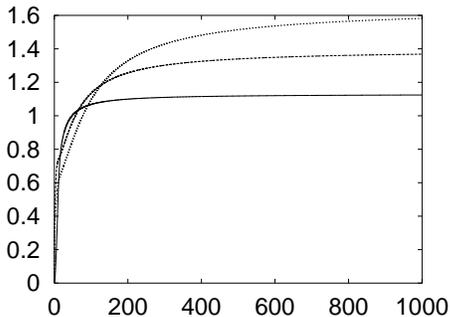}}
\caption{ Initial energy
$E(R)\equiv E(a,0)-E(R,0)$ ($y-$ axis) as a function of $r^*-r^*(a)$ for
$a=2.1$ (solid line), $a=3.1$ (long broken line) and $a=4$ (short broken line).}
\end{figure}

\section{Evolving extremal initial data}

 {\it A. Ringing is loud.} Our primary aim is to find initial
data that give the strongest possible ringing.  
QM  are evidently born by a subclass of multiple backscattering
(see an explanation in \cite{Ching}). 
The  diffusion energy $\delta E_a$ bounds   the energies 
of QM, the tail term and also of the radiation falling to a black hole. 
While we do not have analytic estimates of the shares of the
particular contributing terms in $\delta E_a$, it is obvious that
configurations with large $\kappa $ have some room for robust
oscillations.  For that reason, instead of obeying the commonly
used method of {\it trial and error},  we study GW defined 
by the  {\bf extremal initial data}. 
In the numerical calculations we use the splitting $\Psi =\tilde \Psi +\delta $
(see Sec II.B), since the numerics  is then   more precise.

Fig. 4 presents the radiation corresponding to the $a=2.1$,
$a=3$ and $a=4$ initial pulses as seen by an  "observer"
situated at $r^*_o=280+r^*(a)$.   The $x=0$ point of
the abscissa corresponds to the  moment of time
$t=280$. This train of initial data that moves with the speed of
light is seen earlier ($t<280$)  and it lies to the right
from $x=0$. To the left from $x=0$ we have $t>280$; in the
absence of  the backscattering there would be no signal
at all.    Notice that in the
 cases $a=2.1$ and $a=3.1$ the amplitude of the strongest ringing mode
is about one third of  the  amplitude of the original radiation, as
represented by the function $\tilde \Psi $. The  amplitude of
the strongest ringing  mode decreases with the increase of $a$ -
that is when  the initial pulse is moved  away from the horizon -
but in all cases  it  is of the   order of the initial amplitude.
\begin{figure}[4]
\epsfxsize=6cm
\centerline{\epsffile{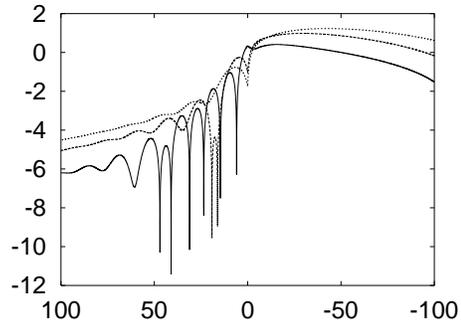}}
\caption{ Temporal dependence of $\ln |\Psi |$
($y-$ axis), as observed at $r^*_0=r^*(a)+280$.
 Solid line,  long broken line and short broken line correspond to $a=2.1$, $a=3.1$ and  $a=4$, respectively. }
\end{figure}

 {\it B. Birth and death of ringing modes.}
QM boil down predominantly in the potential valley $2<R<R_{cr}\approx 3.1$,
where the maximal point $R_{cr}$ is defined by
$\partial_RV|_{R_{cr}}=0$. Their
complex frequencies carry information about the curvature of the
background geometry in this region. This is why
the characteristic features of the ringing modes -
the period  and the damping coefficient of the dominant mode  -
do not depend on initial data. Let us point out, however, that
the amplitudes of the QM depend on the frequency
profile and amplitudes of initial data.
We examined   QM in various observation
points. Fig. 5 (with $a=4$) 
shows clearly that the number of nodes
in a radiation pulse decreases with the distance.
   At the same time  the dominant
oscillations survive and they have the same character (but
their amplitude  can significantly 
increase) while the tail part extends. 

Fig. 5 shows that there are many oscillations
at $r^*_o=r^*(4)+1$, which gradually die when 
the observation point is moved away - only 
one node is seen at $r^*=280$ (Fig. 4). 
This observation that we make
is probably novel in the literature, 
but its explanation  can be standard, within the
scenario described in \cite{Ching}.
 
\begin{figure}[5]
\epsfxsize=6cm
\centerline{\epsffile{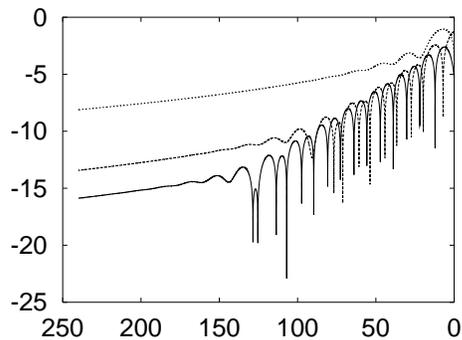}}
\caption{ Backscattered radiation
(values $\ln |\Psi |$ put on $y-$ axis), $a=4$,  as seen by an observer located at $r^*_0=1+r^*(4)$ (solid line), $r_0^*=10+r^*(4)$ (long broken line) and $r_0^*=100+r^*(4)$ (short broken line). }
\end{figure}

\section{Conclusions}

We have found   initial data that correspond to maximal values of
the diffusion parameter $\kappa $; we believe that
these are  
optimal data for having strong QM. The largest amplitudes 
of QM constitute a fraction (c. 1/3) of the largest amplitude 
within   the main pulse of the radiation. 
Our earlier experience \cite{kark2002} 
in numerics suggests that the effect of the
backscatter depends (fixing the distance $a$) mostly on
the frequency, and that if the dominant frequency is low
(as compared to $1/m$),  then the diffusion parametr $\kappa $
does not depend strongly on details of the profile of initial data.
This would imply that in the real collapse the ringing can
be perhaps less vigorous than above, but still of interest,
with    amplitudes smaller by perhaps one order than those
of the main pulse of GW.
Similar conclusions  concerning the strength of QM
  are valid also for the axial quadrupole GW
and for the dipole electromagnetic   waves  \cite{recent}. The
main difference between the   polar and axial
perturbations is that   polar QM   are stronger   and  they seem to be
more "generic", in a sense that will be explained elsewhere \cite{recent}.
The same analysis can be done for higher GW multipoles.

Referring to a specific
case of collapse of BH, we want to stress the following. Our
results   can  be applied to the head-on collision of two spinless
BH.  If one uses  the close approximation limit \cite{Price},  then
the  outgoing part of a radiation that was produced during the
earlier phases, the  merger and the inspiral,  should be
interpreted as our initial data. If these  initial data are
close to "maximal", in a sense used earlier, then the ringing
amplitudes  could be of the order of  GW created during the
coalescence of BH.


\begin{references}
\bibitem{Thorne} C. Cutler et al.,  {\it Phys. Rev. Lett.} {\bf 70}, 2984(1993).
\bibitem{Damour} T. Damour, R. R. Iyer and B. S. Sathyaprakash,
{\it Detecting Binaryy Black Holes With Efficient and Reliable
Templates} in Gravitational Waves - A challenge to Theoretical Physics,
Proceedings of Trieste conference, 2000.
\bibitem{Thorne99} P. R. Brady, J. D. E. Creighton and  K. S. Thorne,
{\it Phys. Rev.} {\bf D58} (1998) 061501.
\bibitem{all} see following review papers and references therein
K. D. Kokkotas and B. Schmidt, {\it Quasi-Normal Modes of
Stars and Black Holes} in Living Reviews in Relativity, published
16 Sept. 1999; H. P. Nollert, {\it Class. Quantum Grav.}, {\bf 16}, 
R159(1999).
\bibitem{Nollert99} H. P. Nollert and R. Price, {\it J. Math. Phys.}
{\bf 40}, 980(1999).
\bibitem{Zerilli} Zerilli, {\it Phys. Rev. Lett.} {\bf }{\bf 24 },
737(1970); {\it Phys. Rev.} {\it \bf D2}, 2141(1970).
\bibitem{energy} The energy density $\rho $ is probably proportional
to the Landau-Lifschitz type energy density of gravitational waves, up
to a four-dimensional divergence term. Here the energy $E(R,t)$
can be regarded as an subsidiary quantity that appears to be useful in
constructing the strongest QM.
\bibitem{Chebyshev} We tried also the Legendre polynomials, $\sin $ and
$\cos $ trigonometric  bases and wavelets, but the numerics was fastest with the
Chebyshev polynomials.
\bibitem{back} There exists a vast literature on the backscatter - see
below and references therein: C. Misner, K. Thorne, J. A. Wheeler,
{\it Gravitation}, Freeman, San Francisco,1973;
 J. Bicak, {\it Gen. Rel. Grav.} {\bf 3}, 331(1972);
R. Price, {\it Phys. Rev} {\bf D5}, 2419(1972);
J. M. Bardeen and W. H. Press, {\it J. Math. Phys.} {\bf 14}, 7(1973);
 B. Mashhoon, {\it Phys. Rev.} {\bf D7}, 2807(1973);
 L. Blanchet and G. Sch\"afer  {\it Class. Quantum Grav.}
{\bf 10}, 2699(1993); E. Malec, {\it Phys. Rev.} {\bf D62}(2000), 084034;
S. Hod, {\it Wave tails in time dependent backgrounds}, gr-qc/0201017.
\bibitem{Gerhard} E. Malec and G. Sch\"afer, 
{\it  Phys.\ Rev.\ } {\bf D64} (2001) 044012.
\bibitem{kark2001} J. Karkowski, E. Malec and Z. \'Swierczy\'nski),  
{\it Acta Phys. Pol.}, {\bf 32}, 3593(2001).
\bibitem{netlib} This can be found in www.netlib.org.
\bibitem{remark} $\Psi_0$ itself   can be set to zero through, say,
introducing a damping factor outside some $R_M>>2m$.
This modification would not  change our conclusions and
we keep a nonzero  asymptotic value  of $\Psi_0$ since
it is more  convenient in doing numerics.
\bibitem{Ching} E. S. C. Ching et al., {\it Phys. Rev. Lett. }
{\bf 74}, 2414(1995); {\it Phys. Rev. } {\bf D52}, 2118(1995).
\bibitem{kark2002} J. Karkowski, E. Malec and Z. \'Swierczy\'nski),
{\it Classical and Quantum Gravity}, {\bf 19}, (2002).
\bibitem{Price} R. H. Price and J. Pullin, {\it Phys. Rev. Lett.}
{\bf 72}, 3297(1994).
\bibitem{recent} J. Karkowski, K. Roszkowski and Z. \'Swierczy\'nski,
work in progress.
\end{references}
\end{document}